\documentclass{llncs}
\newfont{\mycrnotice}{ptmr8t at 7pt}
\newfont{\myconfname}{ptmri8t at 7pt}

\usepackage{caption}
\usepackage{algpseudocode}
\usepackage{amsmath}
\usepackage{amsfonts}
\usepackage{subfigure}
\usepackage{graphicx}

\usepackage{amsfonts}
\usepackage{booktabs}
\usepackage{siunitx}

\usepackage{color}

\usepackage{tikz}

\usepackage{enumitem}

\usepackage{url}
\makeatletter
\def\url@leostyle{%
  \@ifundefined{selectfont}{\def\UrlFont{\sf}}{\def\UrlFont{\small\ttfamily}}}
\makeatother
\newcommand{\eat}[1]{}
\usepackage[linesnumbered,ruled]{algorithm2e}
\usepackage{mdframed} 

\usepackage{xcolor}
\definecolor{light-gray}{gray}{0.9}

\begin{document}

\title{Comparison of Decentralization in DPoS and PoW Blockchains}



\author{Chao Li\inst{1} \and Balaji Palanisamy\inst{2}}
\institute{Beijing Key Laboratory of Security and Privacy in Intelligent Transportation,
Beijing Jiaotong University \\ 
\email{li.chao@bjtu.edu.cn}
\and
School of Computing and Information, University of Pittsburgh \\
\email{bpalan@pitt.edu}
}

\maketitle




\begin{abstract}
Decentralization is a key indicator for the evaluation of public blockchains. 
In the past, there have been very few studies on measuring and comparing the actual level of decentralization between Proof-of-Work (PoW) blockchains and blockchains with other consensus protocols.
This paper presents a new comparison study of the level of decentralization in Bitcoin and Steem, a prominent Delegated-Proof-of-Stake (DPoS) blockchain.
Our study particularly focuses on analysing the power that decides the creators of blocks in the blockchain.
In Bitcoin, miners with higher computational power generate more blocks.
In contrast, blocks in Steem are equally generated by witnesses while witnesses are periodically elected by stakeholders with different voting power weighted by invested stake.
We analyze the process of stake-weighted election of witnesses in DPoS and measure the actual stake invested by each stakeholder in Steem.
We then compute the Shannon entropy of the distribution of computational power among miners in Bitcoin and the distribution of invested stake among stakeholders in Steem.
Our analyses reveal that neither Bitcoin nor Steem is dominantly better than the other with respect to decentralization.
Compared with Steem, Bitcoin tends to be more decentralized among top miners but less decentralized in general.
Our study is designed to provide insights into the current state of the degree of decentralization in DPoS and PoW blockchains.
We believe that the methodologies and findings in this paper can facilitate future studies of decentralization in other blockchain systems employing different consensus protocols.

\end{abstract}

\section{Introduction}

The evolution of Blockchain made Bitcoin the first cryptocurrency that resolves the double-spending problem without the need for a centralized trusted party~\cite{nakamoto2008bitcoin}. 
From then on, rapid advances in blockchain technologies have driven the rise of hundreds of new public blockchains~\cite{kwon2019impossibility}. 
For most of these public blockchains, the degree of decentralization of resources that decide who generates blocks is the key metric for evaluating the blockchain decentralization~\cite{gencer2018decentralization,kwon2019impossibility,wu2019information}. This in turn facilitates further understanding of both security and scalability in a blockchain~\cite{kokoris2018omniledger}.
Intuitively, having only a few parties dominantly possess the resources indicates a more centralized control of blockchain, which is potentially less secure. This is due to the fact collusion among these few parties can be powerful enough to perform denial-of-service attacks against targeted blockchain users and even falsify historical data recorded in blockchain.
More concretely, in a Proof-of-Work (PoW) blockchain such as Bitcoin~\cite{nakamoto2008bitcoin}, a miner possessing higher computational power has a better chance of generating the next block. 
In Bitcoin, a transaction is considered to be `confirmed' after six blocks as it is estimated that the probability of creating a longer fork after six blocks to defeat the one containing the `confirmed' transaction is negligible.
However, the assumption is not held when a few miners possess over half of overall computational power in the network, in which case these miners are able to launch the commonly known 51\% attack to control the blockchain and double-spend any amount of cryptocurrency.
Through selfish mining~\cite{eyal2014majority}, the difficulty of performing 51\% attack could be further reduced to the demand of possessing 33\% of overall computational power, indicating even weaker security in less decentralized blockchains.

Recent research pointed out that Bitcoin shows a trend towards centralization~\cite{beikverdi2015trend,gencer2018decentralization,kwon2019impossibility,tschorsch2016bitcoin,wu2019information}.
For the purpose of reducing the variance of income, the majority of Bitcoin miners have joined large mining pools and a small set of mining pools are now actually controlling the Bitcoin blockchain.
Meanwhile, the use of PoW consensus protocol in Bitcoin requires the decentralized consensus to be made throughout the entire network and the throughput of transactions in Bitcoin is limited by the network scale.
As a result, the 7 transactions/sec throughput in Bitcoin cannot satisfy the need of many practical applications~\cite{croman2016scaling}.
Motivated by these concerns, recently, the Delegated Proof-of-Stake (DPoS) consensus protocol~\cite{larimer2014delegated} is becoming increasingly popular and has given rise to a series of successful blockchains~\cite{kwon2019impossibility,li2019incentivized}. 
The key selling point of DPoS blockchains is their high scalability.
In DPoS blockchains, blocks are generated by a small set of witnesses that are periodically elected by the entire stakeholder community. Thus, the decentralized consensus is only reached among the witnesses and the small scale of the witness set could boost the transaction throughput to support various types of real applications such as a social media platform~\cite{li2019incentivized}.
However, there are disagreements over the level of decentralization in DPoS blockchains.
The supporters believe that, in practice, the design of equally generating blocks among witnesses in DPoS blockchains is less centralized than the current PoW blockchains dominated by few mining pools.
Others believe that, in theory, the very limited scale of the witness set naturally shows a low degree of decentralization.
Existing research works have evaluated the degree of decentralization in Proof-of-Work (PoW) blockchains represented by Bitcoin and Ethereum~\cite{gencer2018decentralization,wu2019information}.
However, there have been very few comparison studies on measuring the actual level of decentralization between PoW blockchains and DPoS blockchains.

This paper presents a comparison study of the level of decentralization in Bitcoin and in Steem~\cite{li2019incentivized}, a prominent DPoS blockchain.
Our  study  particularly  focuses  on  analysing the power that decides the creators of blocks in the blockchain.
We analyze the process of stake-weighted election of witnesses in DPoS and measures the actual stake invested by each stakeholder in Steem.
Similar to the analyses in recent works~\cite{kwon2019impossibility,wu2019information}, we quantify and compare the actual degree of decentralization in these two blockchains by computing the Shannon entropy of the distribution of computational power among miners in Bitcoin and that of the distribution of invested stake among stakeholders in Steem.
Our results shows that the entropy in Steem among top stakeholders is lower than the entropy in Bitcoin among top miners while the entropy in Steem becomes higher than the entropy in Bitcoin when more stakeholders and miners are taken into computation.

Our analyses reveal that neither Bitcoin nor Steem is dominantly better than the other with respect to decentralization. Compared with Steem, Bitcoin tends to be more decentralized among top miners but less decentralized in general.
Our study is designed to provide insights into the current state of the degree of decentralization in representative DPoS and PoW blockchains.
We believe that the methodologies and findings in this paper can facilitate future studies on decentralization in other blockchains and consensus protocols.


The rest of this paper is organized as follows: 
We introduce the background in Section~2.
In Section~3, we present the preliminary measurements for Bitcoin miners and for Steem witnesses.
Then, in Section~4, we present the methodologies of measuring the impact of stakeholders in the process of stake-weighted witness election in Steem.
In Section~5, we quantify and compare the degree of decentralization among Bitcoin miners, among Steem witnesses and among Steem stakeholders.
Finally, we discuss related work in Section~6 and we conclude in Section~7.

\section{Background}

In this section, we introduce the background about the Steem-blockchain~\cite{Steem_blockchain}, including its key application \textit{Steemit}, its implementation of the DPoS consensus protocol and its ecosystem in general.

\begin{figure}[t!]
\centering
{
    \includegraphics[width=8cm,height=6cm]{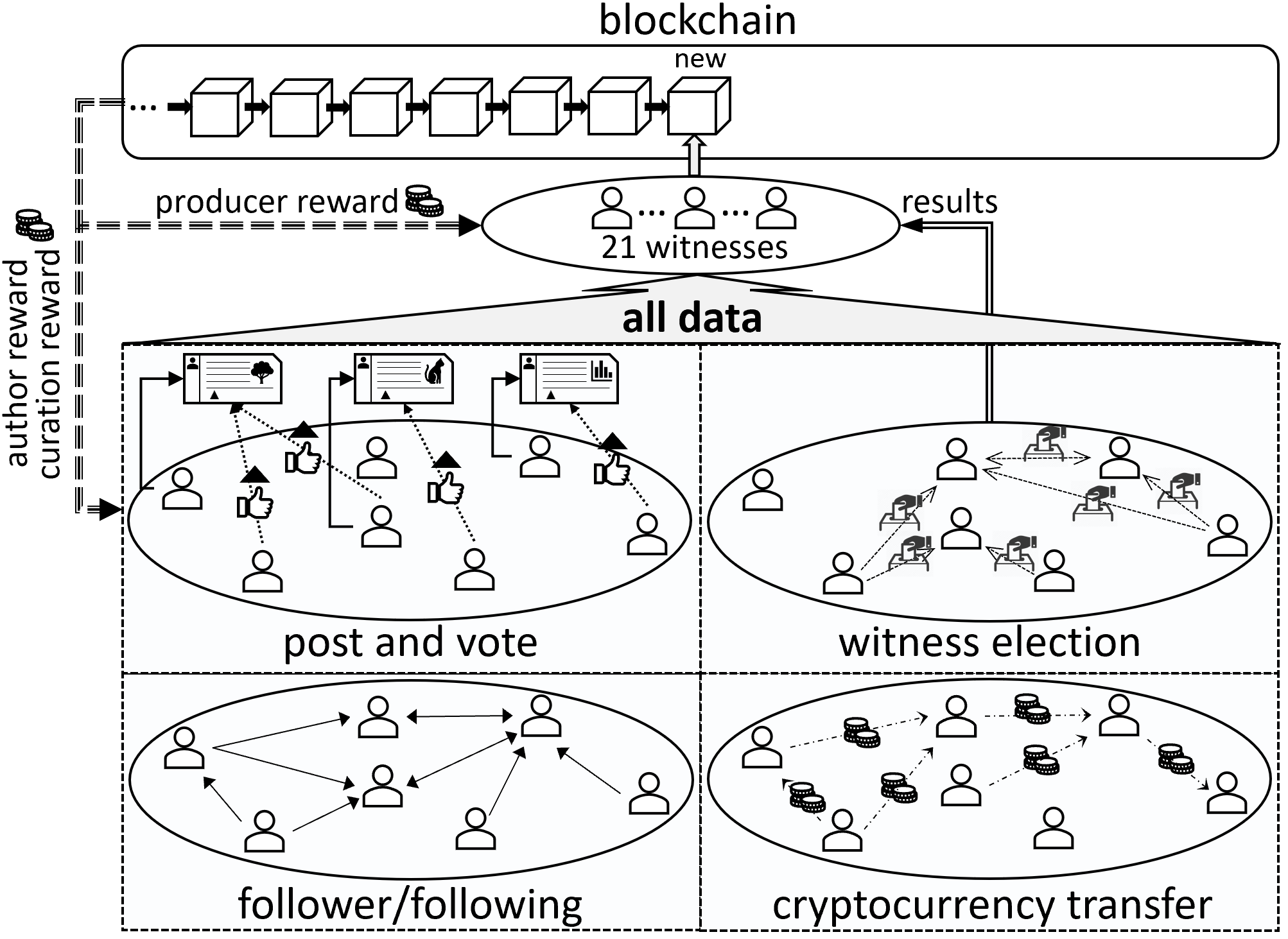}
}
\caption {Steem blockchain overview}
\label{icbc_00} 
\end{figure}

The Steem-blockchain is the backend for \textit{Steemit}, which is the first blockchain-powered social media platform that incentivizes both creator of user-generated content and content curators. \textit{Steemit} has kept its leading position during the last few years and its native cryptocurrency, \textit{STEEM}, has the highest market capitalization among all cryptocurrencies issued by blockchain-based social networking projects. 
Users of \textit{Steemit} can create and share contents as blog posts.
A blog post can get replied, reposted or voted by other users.
Based on the weights of received votes, posts get ranked and the top ranked posts make them to the front page. 
\textit{Steemit} uses the Steem-blockchain to store the underlying data of the platform as a chain of blocks. Every three seconds, a new block is produced, which includes all confirmed operations performed by users during the last three seconds.
\textit{Steemit} allows its users to perform more than thirty different types of operations. In Fig.~\ref{icbc_00}, we display four categories of operations that are most relevant to the analysis presented in this paper. 
While post/vote and follower/following are common features offered by social sites, 
operations such as witness election and cryptocurrency transfer are features specific to blockchains.

Witnesses in \textit{Steemit} are producers of blocks, who continuously collect data from the entire network, bundle data into blocks and append the blocks to the Steem-blockchain.
The role of witnesses in \textit{Steemit} is similar to that of miners in Bitcoin. 
In Bitcoin, miners keep solving Proof-of-Work (PoW) problems and winners have the right to produce blocks.
With PoW, Bitcoin achieves a maximum throughput of 7 transactions/sec ~\cite{croman2016scaling}. 
However, transaction rates of typical mainstream social sites are substantially higher. For example, Twitter has an average throughput of more than 5000 tweets/sec~\cite{Twitter}.
Hence, the Steem blockchain adopts the Delegated Proof of Stake (DPoS)~\cite{larimer2014delegated} consensus protocol to increase the speed and scalability of the platform without compromising the decentralized reward system of the blockchain.
In DPoS systems, users vote to elect a number of witnesses as their delegates. In \textit{Steemit}, each user can vote for at most 30 witnesses. The top-20 elected witnesses and a seat randomly assigned out of the top-20 witnesses produce the blocks.
With DPoS, consensus only needs to be reached among the  21-member witness group, rather than the entire blockchain network like Bitcoin, which significantly improves the system throughput.

We now present the process of stake-weighted witness election with more details.
Any user in \textit{Steemit} can run a server, install the Steem-blockchain and synchronize the blockchain data to the latest block.
Then, by sending a \textit{witness\_update} operation to the network, the user can become a witness and have a chance to operate the website and earn producer rewards if he or she can gather enough support from the electors to join the 21-member witness group.
A user has two ways to vote for witnesses. 
The first option is to perform \textit{witness\_vote} operations to directly vote for at most 30 witnesses.
The second option is to perform a \textit{witness\_proxy} operation to set another user as an election proxy.
The weight of a vote is the sum of the voter's own stake and the stake owned by other users who have set the voter as proxy.
For example, Alice may set Bob to be her proxy. Then, if both Alice and Bob own \$100 worth of stake, any vote cast by Bob will be associated with a weight of \$200 worth of stake. Once Alice deletes the proxy, the weight of Bob's votes will reduce to \$100 worth of stake immediately.

The ecosystem in Steem is a bit complex.
Like most blockchains, the Steem-blockchain issues its native cryptocurrencies called \textit{STEEM} and Steem Dollars (\textit{SBD}).
To own stake in \textit{Steemit}, a user needs to `lock' \textit{STEEM}/\textit{SBD} in \textit{Steemit} to receive Steem Power (\textit{SP}) at the rate of $1\ STEEM = 1\ SP$ and each $SP$ is assigned about 2000 vested shares (\textit{VESTS}) of \textit{Steemit}.
A user may withdraw invested \textit{STEEM}/\textit{SBD} at any time, but the claimed fund will be automatically split into thirteen equal portions to be withdrawn in the next thirteen subsequent weeks.
For example, in day 1, Alice may invest 13 \textit{STEEM} to \textit{Steemit} that makes her vote obtain a weight of 13 \textit{SP} (about 26000 \textit{VESTS}). Later, in day 8, Alice may decide to withdraw her 13 invested \textit{STEEM}. Here, instead of seeing her 13 \textit{STEEM} in wallet immediately, her \textit{STEEM} balance will increase by 1 \textit{STEEM} each week from day 8 and during that period, her \textit{SP} will decrease by 1 \textit{SP} every week.
In the rest of this paper, for ease of exposition and comparison, we transfer all values of \textit{STEEM/SBD/SP/VESTS} to VESTS, namely the stake in Steem, based on $1\ SBD \approx 0.4\ STEEM = 0.4\ SP \approx 800\ VESTS$~\cite{SteemPrice}.

\section{Data collection and preliminary measurements}

In this section, we describe our data collection methodology and present some preliminary measurements among Bitcoin miners and among Steem witnesses.

\subsection{Data collection}
\label{sec3.1}
The Steem-blockchain offers an Interactive Application Programming Interface (API) for developers and researchers to collect and parse the blockchain data~\cite{SteemAPI}.
From block 24,671,073 to block 25,563,499, we collected 892,426 Steem blocks produced during a time period of one month.
The Bitcoin blockchain offers similar API at~\cite{BitcoinAPI}.
From block 534,762 to block 539,261, we collected 4,499 Bitcoin blocks produced during the same one-month time period. 
We have recently released a dataset for \textit{Steemit} named \textit{SteemOps} at: \\ 
\centerline{\url{https://github.com/archerlclclc/SteemOps}}

\subsection{Preliminary measurements}

\begin{figure}[t!]
\centering
{
    \includegraphics[width=1\columnwidth]{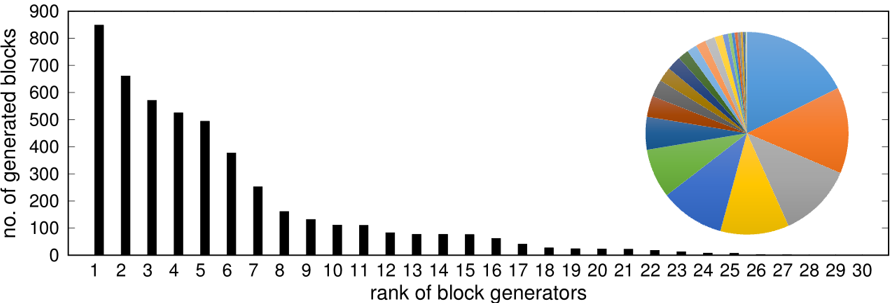}
}
\caption {Distribution of blocks generated by top-30 miners in Bitcoin}
\label{icbc_01} 
\end{figure}

\begin{figure}[t!]
\centering
{
    \includegraphics[width=1\columnwidth]{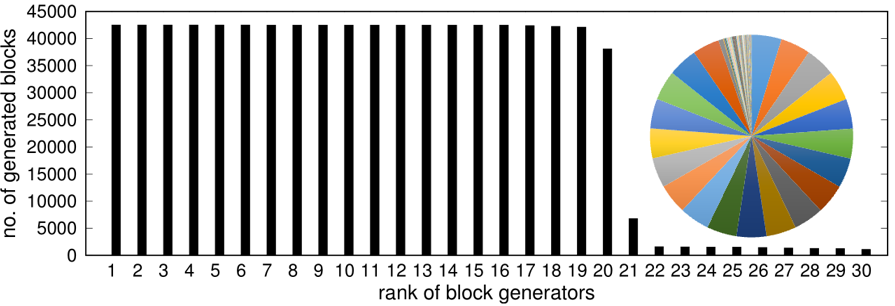}
}
\caption {Distribution of blocks generated by top-30 witnesses in Steem}
\label{icbc_02} 
\end{figure}

To understand the degree of decentralization in Bitcoin and Steem, we start measurements by parsing the collected blocks and counting the number of blocks produced by each generator, namely each Bitcoin miner or Steem witness.
The results of distributions of blocks created by top-30 generators in Bitcoin and Steem are shown in Fig.~\ref{icbc_01} and Fig.~\ref{icbc_02}, respectively.

We notice that 4,430 out of 4,499 Bitcoin blocks (98.6\%) were generated by mining pools, which illustrates the domination of mining pools in current Bitcoin mining competitions.
The top 5 mining pools mined 848 (18.9\%), 661 (14.7\%), 571 (12.7\%), 525 (11.7\%) and 494 (11.0\%) blocks, respectively.
As revealed by the pie chart plotted in Fig.~\ref{icbc_01}, the sum of blocks produced by the top 4 mining pools, namely 2605 (57.9\%) blocks in total, has already exceeded the bar of launching 51\% attack.
The top 2 mining pools generated 1,509 (33.5\%) blocks have been higher than the 33\% bar suggested by the selfish mining research~\cite{eyal2014majority}.
Intuitively, we see how fragile the security in the current Bitcoin blockchain is in practice.
It is true that mining pools may not want to attack Bitcoin as it would decrease the market price of Bitcoin and damage the profit of themselves.
However, the risk is still high because compromising two mining pools is much easier than attacking a large distributed network and it is also hard to quantify and compare the short-term profit of performing double-spend before Bitcoin gets crashed and the long-term profit of keeping honest.

Next, we notice that 844,390 out of 892,426 Steem blocks (94.6\%) were generated by the top-20 witnesses.
Interestingly, as indicated by the pie chart plotted in Fig.~\ref{icbc_02}, the mean and standard deviation of blocks generated by top 20 witnesses are 42,219 and 978 respectively, indicating that these top 20 witnesses continuously reserved 20 seats in the 21-member witness group and generated nearly the same amount of blocks in the selected month.
Although we do not see that the seats were frequently switched among a larger set of witnesses, intuitively, the distribution of blocks among Steem witnesses shows a trend of a higher degree of decentralization.
However, unlike mining pools in Bitcoin that possess computational power directly determining the number of blocks they could mine, witnesses in Steem do not own the similar resources themselves because their sears in the 21-member witness group are determined by the VESTS (i.e., stake) accumulated from the votes cast by stakeholders in the entire network.
That is, block generation in Steem and other DPoS blockchains is actually controlled by stakeholders who possess the power of appointing and removing witnesses at any time.
Therefore, 
in the next section, we investigate the impact of stakeholders in the process of stake-weighted witness election in Steem, which would facilitate the quantification of the actual level of decentralization in Steem.

\section{Measurements on impact of stakeholders in Steem}

In this section, we measure the actual impact of stakeholders in determining active witnesses in Steem, namely in actually controlling the Steem blockchain. 
We first present our measuring methodology and then the results.

\subsection{Methodology}

Each stakeholder could cast at most 30 votes and each vote is weighted by net VESTS, namely the sum of the stakeholder's pure VESTS and the VESTS received from other stakeholders who have set this stakeholder as proxy.
Therefore, we first investigate pure VESTS that each stakeholder possesses and compute net VESTS belonged to each stakeholder in witness election by combining pure VESTS with VESTS received from other stakeholders.

However, computing the degree of decentralization in Steem based on net VESTS is inaccurate for two reasons:
(1) stakeholders may not choose to vote for any witness and therefore, VESTS owned by this type of stakeholders has no contribution to witness election and these stakeholders have no actual control of the blockchain;
(2) a stakeholder can cast at most 30 votes, but not all stakeholders cast 30 votes in full.
For example, Alice may cast 20 votes to 20 witnesses and Bob may cast a single vote to a single witness.
Assuming that Alice and Bob have the same net VESTS and all these voted witnesses generated the same amount of blocks, we consider that the impact of Alice in witness election is twenty times that of Bob.
In other words, what determines the actual impact of a stakeholder in witness election is the accumulated VESTS from all his/her votes, namely the multiplication of net VESTS and the number of votes.

To do that, we need to collect the number of votes cast by each stakeholder.
After that, we compute accumulated VESTS from votes cast by each stakeholder and accumulated VESTS from votes received by each witness.
Finally, we re-allocate the 892,426 Steem blocks to the stakeholders based on accumulated VESTS of their votes and we investigate the distribution of blocks among stakeholders after the re-allocation.

\subsection{Measurement results}

\begin{figure}[t!]
\centering
{
    \includegraphics[width=1\columnwidth]{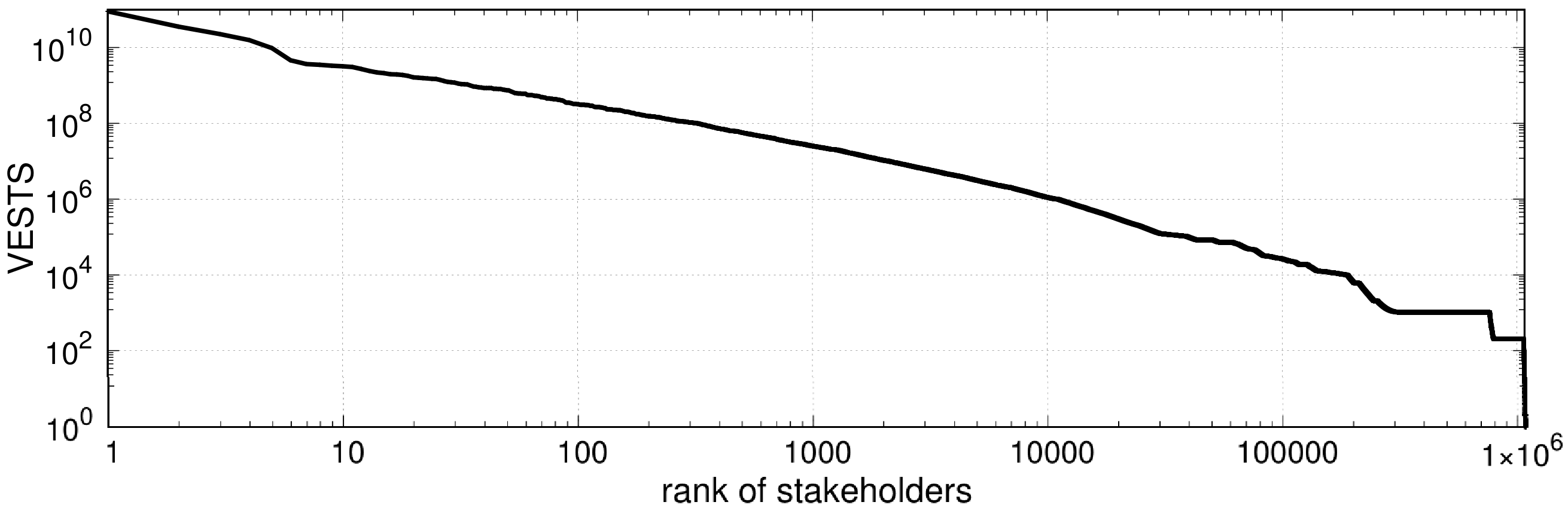}
}
\caption {Distribution of pure VESTS among stakeholders with at least 1 VESTS}
\label{icbc_03} 
\end{figure}

Following the above-mentioned methodology, the measurements start by investigating pure VESTS that each stakeholder has in Steem at the moment of block 25,563,499.
Fig.~\ref{icbc_03} shows pure VESTS of 1,077,405 stakeholders who have at least 1 VESTS.
Here, stakeholders are sorted based on their pure VESTS.
The results indicate a heavy-tailed distribution.
We find that the top-10 stakeholders possess 1.93E+11 VESTS, about 48.5\% of overall VESTS.
In contrast, the last 1,000,000 stakeholders possess only 4.83E+09 VESTS, about 1.2\% of overall VESTS.
The top-1, top-2 and top-5 stakeholders possess 9.00E+10 (22.7\%), 1.26E+11 (37.5\%) and 1.74E+11 (43.9\%) VESTS, respectively.
Therefore, the results here suggest a trend towards centralization in Steem, which is similar to that in Bitcoin.

\begin{figure}[t!]
\centering
{
    \includegraphics[width=1\columnwidth]{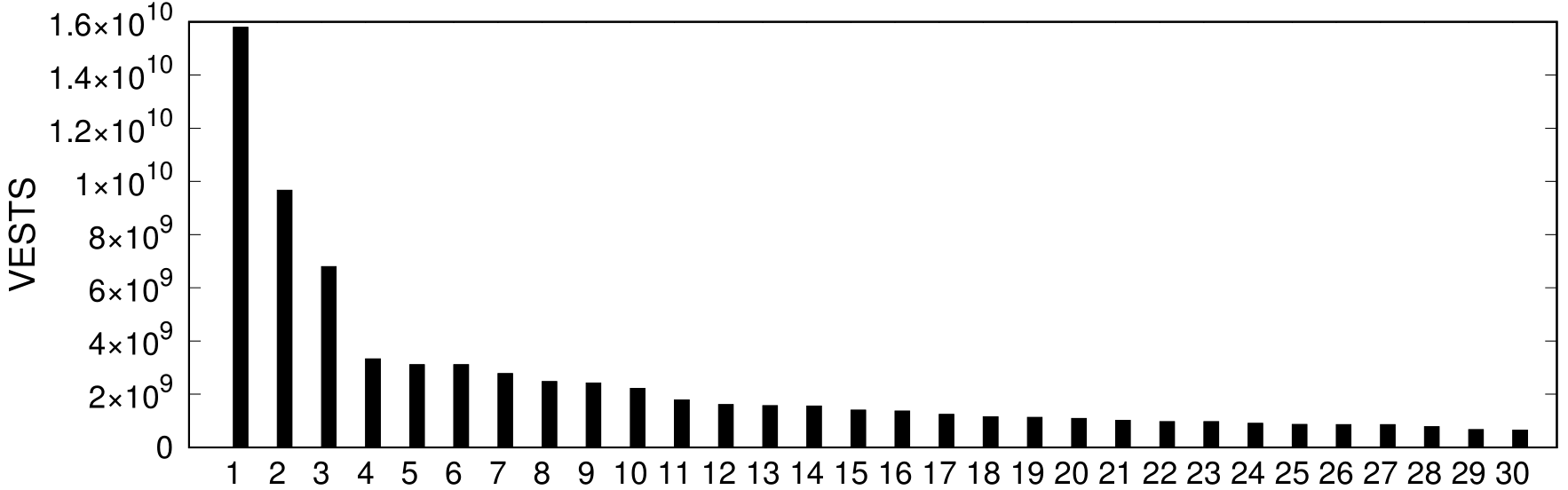}
}
\caption {Distribution of net VESTS among top-30 stakeholders}
\label{icbc_04} 
\end{figure}

Next, Fig.~\ref{icbc_04} presents the distribution of net VESTS among top-30 stakeholders who cast at least one vote and also set no proxy at the moment of block 25,563,499.
Stakeholders here are ranked based on their net VESTS, namely the sum of pure VESTS and received VESTS.
In the rest of this paper, for the ease of presentation, we use the same rank for stakeholders and the phrase 'top-30 stakeholders' refers to the top-30 stakeholders sorted by net VESTS.
The net VESTS belonging to top-1, top-3 and top-5 stakeholders are 1.58E+10, 3.23E+10 and 3.87E+10 respectively, which are only 17.6\%, 21.7\% and 22.2\% of the pure VESTS possessed by stakeholders with the same ranks.
The main reason is that many stakeholders neither cast a single vote nor set other stakeholders as their proxy.
Another interesting observation is that only seven stakeholder possessing top-30 pure VESTS also own top-30 net VESTS.
One possible reason is that many stakeholders may want to isolate operations about witness election from other types of operations, so they create a new account to be their proxy for voting witnesses.

\begin{figure}[t!]
\centering
{
    \includegraphics[width=1\columnwidth]{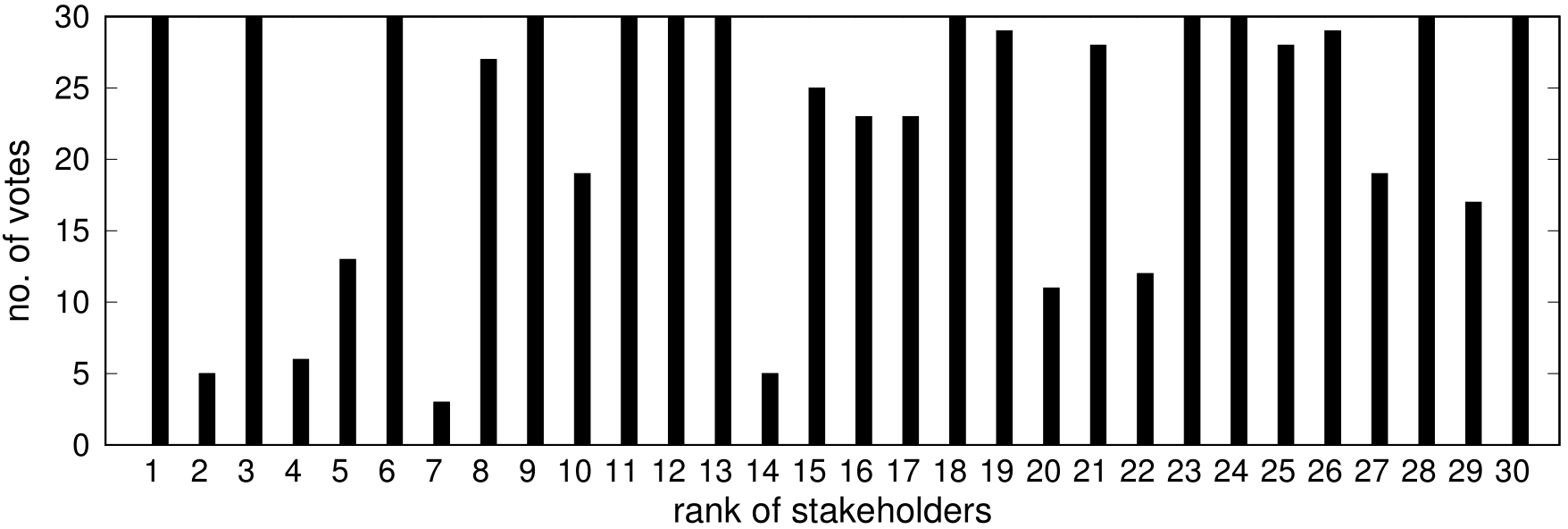}
}
\caption {No. of votes casted by top-30 stakeholders}
\label{icbc_05} 
\end{figure}

After measuring both pure VESTS and net VESTS, we plot the number of votes cast by top-30 stakeholders ranked by net VESTS in Fig.~\ref{icbc_05}.
The results show that only 12 out of top-30 stakeholders have cast 30 votes in full.
The number of votes is quite a personal option. 
The Steem team encourages the community to actively participate in witness election, but some stakeholders may have a strict standard in mind and believe in only less than 30 witnesses, so they do not aim at maximizing their impact in witness election by casting exact 30 votes.

\begin{figure}[t!]
\centering
{
    \includegraphics[width=1\columnwidth]{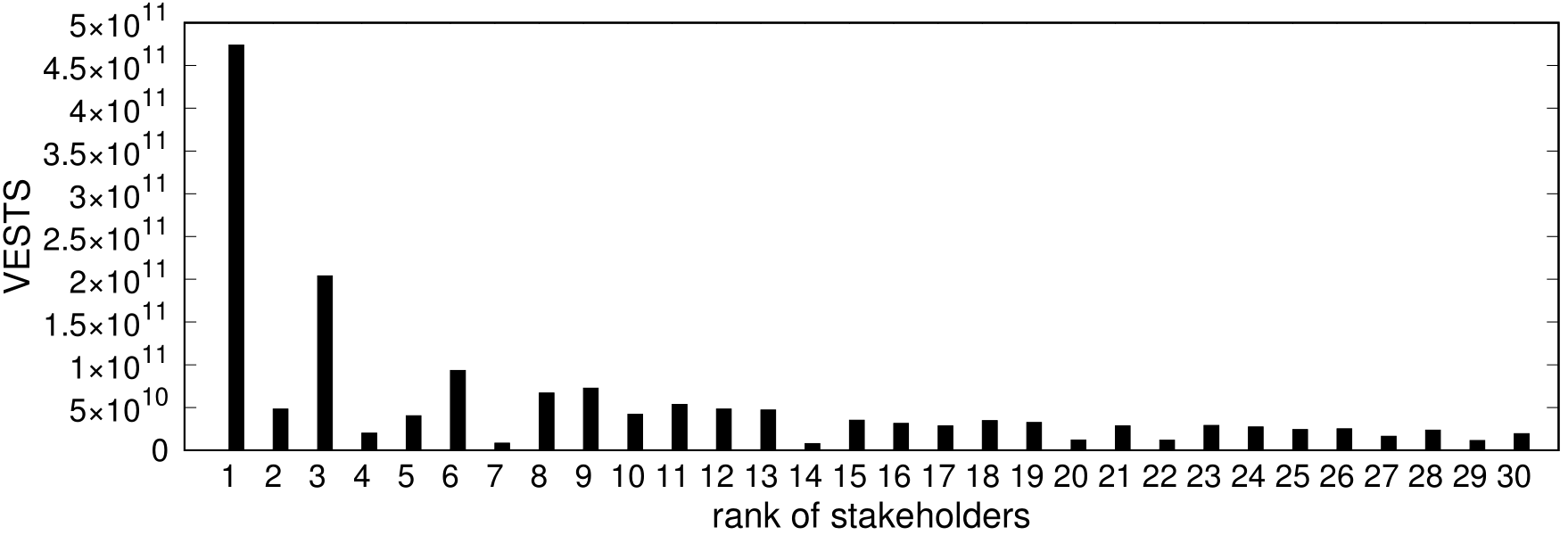}
}
\vspace{-3mm}
\caption {Accumulated VESTS from votes cast by top-30 stakeholder}
\label{icbc_06} 
\end{figure}

We can now compute accumulated VESTS from votes cast by top-30 stakeholders by multiplying net VESTS and the number of cast votes.
The results are shown in Fig.~\ref{icbc_06}.
As can be seen, the 12 stakeholders who cast 30 votes in full amplify their net VESTS by a factor of 30 and maximize their impact in witness election.
In contrast, some stakeholders (i.e, rank-2, rank-4 and rank-7), who cast a few votes, amplify their net VESTS by a much smaller factor, which actually reduces their impact in witness election.
It is hard to comment whether the strategy of always casting 30 votes in full is healthy for Steem or not, but stakeholders who intend to maximize their control in the Steem blockchain would be incentivized to always follow this strategy and their impact in Steem may eventually overtake the impact of stakeholders who cast a personalized number of votes.

\begin{figure}[t!]
\centering
{
    \includegraphics[width=1\columnwidth]{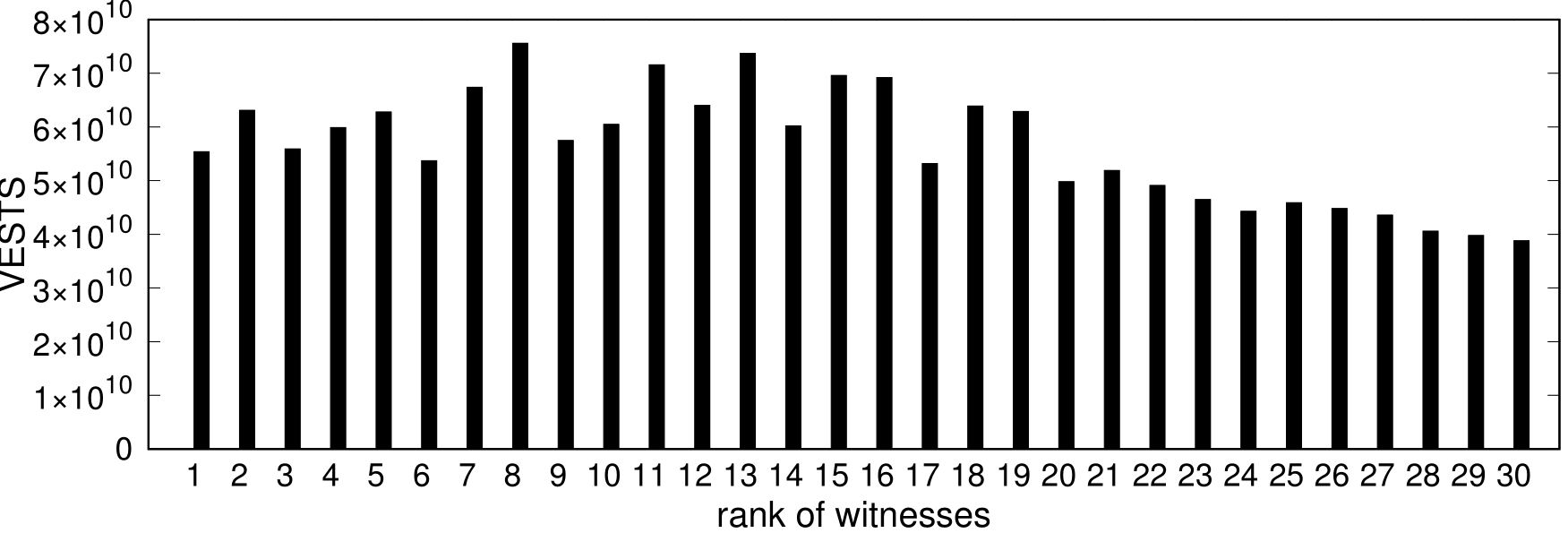}
}
\vspace{-3mm}
\caption {Accumulated VESTS from votes received by top-30 witness}
\label{icbc_07} 
\end{figure}

Following the same methodology, we can also compute accumulated VESTS from votes received by each witness, namely the sum of net VESTS of each vote received by a witness.
The results are shown in Fig.~\ref{icbc_07}.
The mean and standard deviation of accumulated VESTS received by top-30 witnesses are 5.65E+10 and 1.06E+10, respectively.
The comparison between Fig.~\ref{icbc_07} and Fig.~\ref{icbc_06} thus tells us that the degree of decentralization among VESTS accumulated via witnesses tends to be much higher than that among VESTS accumulated via stakeholders.
The relationship between stakeholders and witnesses is similar to that between shareholders and managers in a corporation.
A corporation is actually controlled by its shareholders, especially the major ones, and likewise the Steem blockchain is actually controlled by shareholders, especially the ones possessing large amounts of net VESTS.
Therefore, we consider that the actual degree of decentralization in Steem is expressed through that among stakeholders, rather than that among witnesses.
Another interesting observation is that the amount of blocks generated by a witness is not exactly proportional to its accumulated VESTS.
For instance, the rank-8 witness in Fig.~\ref{icbc_07} received the most VESTS from stakeholders.
This scenario may be caused by the bias that we took the pure VESTS and votes as a snapshot at block 25,563,499 while some stakeholders may have changed their pure VESTS or votes during the one-month time period.

\begin{figure}[t!]
\centering
{
    \includegraphics[width=1\columnwidth]{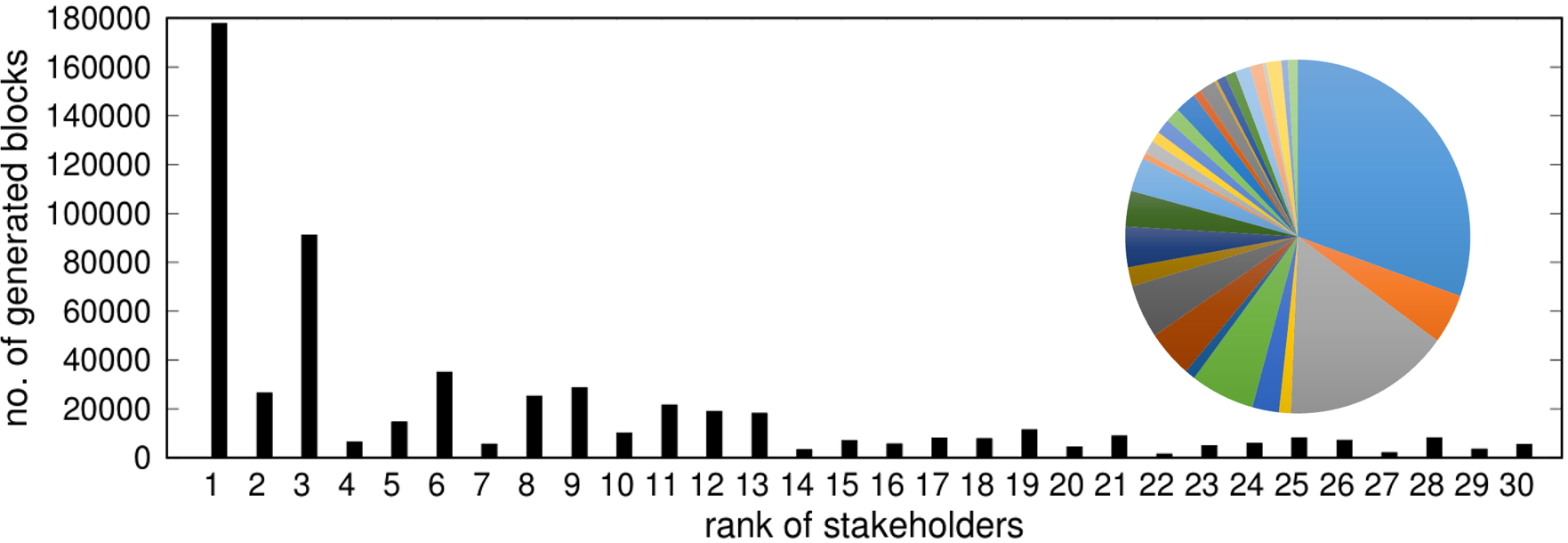}
}
\caption {Distribution of blocks re-allocated to top-30 stakeholders}
\label{icbc_09} 
\end{figure}

Finally, we can observe the actual degree of decentralization in Steem by re-allocating the 892,426 Steem blocks to the stakeholders based on accumulated VESTS of their votes.
The results are shown in Fig.~\ref{icbc_09}, which is obviously more skewed than the distribution we see in Fig.~\ref{icbc_02}.
From the pie chart, we can see that the majority of the 892,426 Steem blocks have been re-allocated to only a few major stakeholders. 
The amounts of blocks re-allocated to the rank-1 stakeholder and to rank-3 stakeholder are 177,698 and 
91,207 respectively, namely 30.5\% and 16.1\% of blocks re-allocated to top-30 stakeholders or 20.0\% and 10.1\% of overall blocks.

We have measured the distribution of blocks among Bitcoin miners (Fig.~\ref{icbc_01}), among Steem witnesses (Fig.~\ref{icbc_02}) and finally among Steem stakeholders (Fig.~\ref{icbc_09}). 
In the next section, we will quantitatively analyze the degree of decentralization in Bitcoin and Steem.

\section{Quantitative analysis of the degree of decentralization}

In this section, similar to the analyses in recent works~\cite{kwon2019impossibility,wu2019information}, we quantify and compare the degree of decentralization in Bitcoin and Steem by computing the Shannon entropy~\cite{shannon1948mathematical}.
We first normalize the three distributions from Fig.~\ref{icbc_01}, Fig.~\ref{icbc_02} and Fig.~\ref{icbc_09} to display a more intuitive comparison among the three distributions.
We then quantify the Shannon entropy (or entropy for short) of the three distributions.

\begin{figure}[t!]
\centering
{
    \includegraphics[width=0.8\columnwidth]{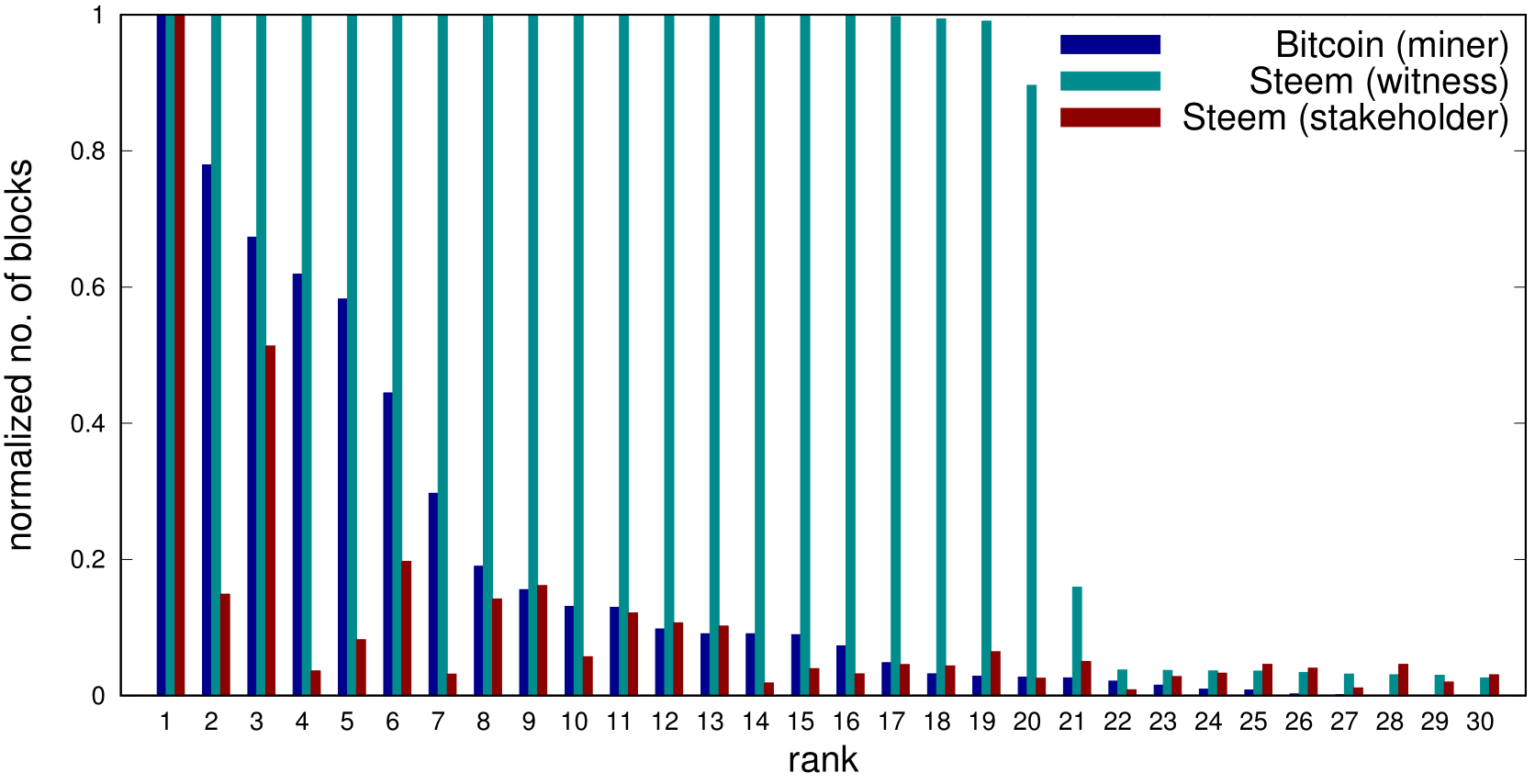}
}
\caption {Comparison of normalized distributions}
\label{icbc_10} 
\end{figure}

\subsection{Normalized comparison}
\label{s5.1}
We display the three normalized distributions at Fig.~\ref{icbc_10}.
Specifically, the normalized number of blocks at $y$-axis has the range [0,1], where the upper bound indicates the number of blocks generated by the rank-1 Bitcoin miner in the Bitcoin (miner) distribution at Fig.~\ref{icbc_01}, the amount of blocks generated by the rank-1 Steem witness in the Steem (witness) distribution at Fig.~\ref{icbc_02} and the amount of blocks re-allocated to the rank-1 Steem stakeholder in the Steem (stakeholder) distribution at Fig.~\ref{icbc_09}.

\noindent \textbf{Bitcoin (miner) vs. Steem (witness):}
The comparison between Bitcoin miners and Steem witnesses seems to support the opinion of supporters of Steem, who believe that the design of equally generating blocks among witnesses in DPoS blockchains is less centralized than the current PoW blockchains dominated by few mining pools.
It is true that the distribution of Bitcoin miners is quite skewed while the distribution of Steem witnesses is quite flat before rank 20.
It is also true that the miners/witnesses after rank 20 generate quite limited amounts of blocks.
However, the fundamental problem of this opinion is that measuring witnesses is not the proper approach for understanding the actual degree of decentralization in Steem.
In Bitcoin, miners possess computational power that determines their probability of mining blocks.
In Steem, witnesses do not possess VESTS that determines their chance to join the 21-member witness group.
As a result, collusion among the few top stakeholders can easily remove the majority of current top witnesses out of the 21-member witness group and control the seats in their hands.
Therefore, to understand the actual situation, we need to compare Bitcoin miners with Steem stakeholders.

\noindent \textbf{Bitcoin (miner) vs. Steem (stakeholder):}
The comparison between Bitcoin miners and Steem stakeholders can be analyzed in three ranges.
From rank 1 to 7, we see that only three out of the top 7 stakeholders cast 30 votes in full, which indicates a significant advantage owned by these three stakeholders in controlling Steem blockchain.
We also see that the distribution among top-7 stakeholders is much skewed than that among top-7 Bitcoin miners.
From rank 8 to 13, the figure suggests no significant advantage between the two distributions.
After rank 13, we see that the distribution of Bitcoin miners quickly drops to zero while the distribution of Steem stakeholders keeps a flat shape.
Overall, intuitively, the degree of decentralization in Bitcoin seems to be higher among top miners than that among top Steem stakeholders but be lower among low ranking miners than that among low ranking Steem stakeholders.

\begin{figure}[t!]
\centering
{
    \includegraphics[width=0.6\columnwidth]{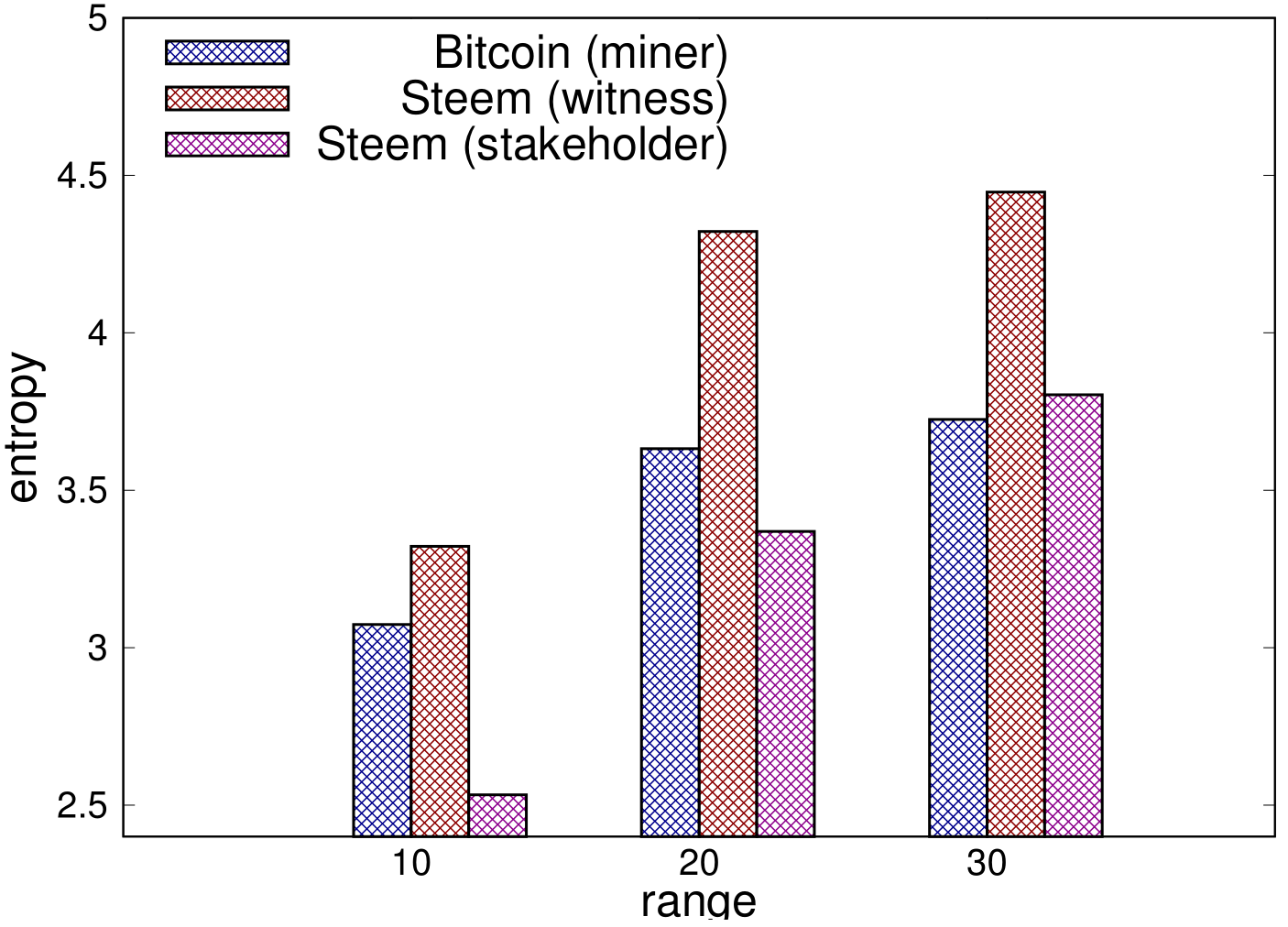}
}
\caption {Comparison of entropy}
\label{icbc_11} 
\end{figure}

\subsection{Quantitative analysis}

We compute the Shannon entropy $entropy$ of the three distribution via:
$$p_i = \frac{b_i}{\sum_{i=1}^{r} b_i}$$
$$entropy = - \sum_{i=1}^{r} p_i \log_2 p_i$$
where $b_i$ denotes the amount of blocks generated by miners/witnesses or re-allocated to stakeholders and $r$ denotes the range of miners/witnesses/stakeholders that entropy is computed for.
For instance, $r=10$ indicates that entropy is computed based on blocks belonging to top-10 miners/witnesses/stakeholders.

We show the computed values of entropy for $r=10,20,30$ in Fig.~\ref{icbc_11}.
We can see that the results well support our analysis in Section~\ref{s5.1}.
First, the entropy among witnesses in all selected ranges keeps being the highest one, which suggests a higher degree of decentralization among witnesses because a group of witnesses has a close probability to be the generator of a block and the uncertainty of inferring the generator is quite high.
Next, the entropy among Bitcoin miners is higher than that among Steem stakeholders in ranges $r=10$ and $r=20$ but turns to become lower in the range $r=30$.
The reason is that the impact of the few major Steem stakeholders is more significant in the sets of top-10 and top-20 stakeholders but is less significant when the computation counts more low ranking stakeholders.
The results thus indicate that, compared with Bitcoin, Steem tends to be more centralized among top stakeholders but more decentralized in general.

\section{Related work}

Most of related works on decentralization in blockchains have focused on Bitcoin~\cite{beikverdi2015trend,eyal2015miner,eyal2014majority,tschorsch2016bitcoin}. 
These works pointed out that Bitcoin shows a trend towards centralization because of the emergence of mining pools.
In~\cite{eyal2014majority}, authors proposed the selfish mining, which reduces the bar of performing 51\% attack to possessing over 33\% of computational power in Bitcoin.
Later, authors in~\cite{eyal2015miner} analyzed the mining competitions among mining pools in Bitcoin from the perspective of game theory and proposed that a rational mining pool may get incentivized to launch a block withholding attack to another mining pool.
Besides Bitcoin, recent work has analyzed the degree of decentralization in Steem~\cite{li2019incentivized}. The work analyzed the process of witness election in Steem from the perspective of network analysis and concluded that the Steem network was showing a relatively low level of decentralization.

Recently, there have been a few studies on comparing the level of decentralization between different blockchains~\cite{gencer2018decentralization,kwon2019impossibility,wu2019information}.
Specifically, the work in~\cite{gencer2018decentralization} compared the degree of decentralization between Bitcoin and Ethereum and concluded that neither Bitcoin nor Ethereum was performing strictly better properties than the other.
The work in~\cite{wu2019information} also focused on investigating Bitcoin and Ethereum, but by quantifying the degree of decentralization with Shannon entropy. Their results indicate that Bitcoin tends to be more decentralized than Ethereum.
The closest work to our study in this paper is the study in~\cite{kwon2019impossibility}, where authors analyzed the degree of decentralization in dozens of blockchains including Bitcoin and Steem.
However, the degree of decentralization in Steem in~\cite{kwon2019impossibility} was computed among witnesses rather than stakeholders, which in our opinion fails to reflect the actual degree of decentralization in a DPoS blockchain.
To the best of our knowledge, our paper is the first research work that quantifies the degree of decentralization in a DPoS blockchain from the perspective of stakeholders after careful analysis and measurements of the witness election.

\section{Conclusion}
In this paper, we present a comparison study of the degree of decentralization in PoW-powered Bitcoin blockchain and DPoS-powered Steem blockchain.
Our study analyzes the process of stake-weighted election of witnesses in the DPoS consensus protocol and measures the actual stake invested by each stakeholder in Steem.
We then quantify and compare the actual degree of decentralization in the two blockchains by computing the Shannon entropy.
Our measurements indicate that, compared with Steem, Bitcoin tends to be more decentralized among top miners but less decentralized in general.
We believe that the methodologies and findings in this paper can facilitates future studies on decentralization in other blockchains and consensus protocols.


\renewcommand\refname{Reference}

\bibliographystyle{plain}
\urlstyle{same}

\bibliography{main.bib}

\end{document}